\documentclass[aps,pra,twocolumn,groupedaddress]{revtex4}
\usepackage{graphicx}

\begin{document}
\title{Properties of quasi two-dimensional condensates in highly anisotropic traps}
\author{G. Hechenblaikner, J. M. Krueger, and C. J. Foot}

\affiliation{Clarendon Laboratory, Department of Physics, University of Oxford,\\
Parks Road, Oxford, OX1 3PU, \\
United Kingdom.}
\date{\today}

\begin{abstract}
We theoretically investigate some of the observable properties of
quasi two-dimensional condensates. Using a variational model based
on a Gaussian-parabolic trial wavefunction we calculate chemical
potential, condensate size in time-of-flight, release energy and
collective excitation spectrum for varying trap geometries and
atom numbers and find good agreement with recent published
experimental results.
\end{abstract}

\maketitle

\section{Introduction}

Bose-Einstein condensation of dilute atomic gases has been
achieved in a variety of magnetic and optical dipole force traps
with different geometries. There is considerable interest in
studying the properties of these ultra-cold gases under conditions
where the confinement gives a system with dimensionality less than
3; recent experiments in optical lattices have observed the
properties of a one-dimensional Tonks gas \cite{Paredes2004a} in
which bosons show fermionic properties. Several other experiments
have realised conditions of two-dimensional confinement, i.e.
where one degree of freedom is frozen out \cite{Gorlitz2001},
\cite{Rychtarik2004}, \cite{Smith2004b}; however, the new physics
in this regime remains to be explored: a two dimensional Bose gas
in a homogenous potential does not undergo Bose-Einstein
condensation (BEC), instead there is a
Berenzinskii-Kosterlitz-Thouless transition (a topological phase
transition mediated by the spontaneous formation of vortex pairs),
a system that is superfluid even though it does not possess
long-range order.
 This is counter the usual picture of
superfluidity in three dimensions explained in terms of a
macroscopic wavefunction describing the whole system. Early
experiments on the KT transition were carried out with films of
superfluid $^4\text{He}$ \cite{Bishop1978}\cite{Kosterlitz1973}
and more recent ones include the observation of quasi-condensates
in thin layers of spin-polarized hydrogen \cite{Safonov1998}.

In recent experiments, BECs created in conventional
three-dimensional magnetic traps have been put into the quasi-2d
(Q2D) regime through the addition of an optical potential. In this
limit the interaction energy, proportional to the chemical
potential $\mu$, is on the order of or smaller than the harmonic
oscillator level spacing and along the tightly confined axial
direction the characteristics of the condensate are those of an
ideal gas . The crossover to the Q2D regime was first observed in
\cite{Gorlitz2001} and in \cite{Rychtarik2004}  by continuously
removing atoms from a highly anisotropic trap to decrease the
interaction energy. In the experiment described in
\cite{Smith2004b} the Q2D crossover is observed by gradually
increasing  the trap anisotropy from moderate to very large values
whilst keeping the number of atoms fixed.

The main aim of this paper is to use a simple theoretical model to
examine some of the characteristic properties of Q2D condensates,
e.g. the chemical potential, release energy and quadrupole mode
spectra are each calculated for a variety of various trap
anisotropies which are achieved in the experiments. Note that a
similar variational approach was taken in \cite{Das2002a} which
investigates the crossover to lower dimensions in general. We
extend and expand upon this analysis by making a more general
ansatz for the trial wavefunction (breaking the axial symmetry)
and deriving analytical expressions for chemical potential, total
energy, release energy and radial condensate widths in the Q2D
regime; a polynomial equation the solution of which gives the
axial and radial condensate sizes in all regimes is also derived.
It is used to calculate physical properties in all regimes and
plot the crossover from the hydrodynamic to the Q2D limit for a
number of physical quantities. The theoretical results are in very
good agreement with the data of a recent experiment in
\cite{Smith2004b}. We numerically calculate the frequency spectrum
for the lowest quadrupole modes and also find an approximate
analytic expression.

The paper is structured as follows:  In the second section we
develop a formalism that allows us to calculate the ground state
and the dynamics of a Q2D gas; starting from a Gaussian-parabolic
trial wavefunction we use a variational method to obtain a set of
differential equations for the dynamics of the Q2D condensate and
a polynomial equation for the ground state. We then examine the
criteria for Q2D and calculate the chemical potential and energy
per particle across the hydrodynamic-Q2D crossover. In section IV
we calculate the size and release energy of condensates in
time-of-flight which are released from traps of varying anisotropy
and make a quantitative comparison with recent experimental data
\cite{Smith2004b}. The final section examines the change of
collective excitation frequencies across the Q2D transition for
the two quadrupole modes with zero angular momentum followed by
the conclusions and a brief outlook.

\section{The hybrid variational model}
Condensates are usually trapped in harmonic potentials given by
\begin{equation}
V_{ext}=\frac{m}{2}\omega_0^2\sum_i\lambda_i^2 x_i^2,
\label{globalpot}
\end{equation}
where the $\lambda_i(t)$ denote the trap anisotropies which can in
general depend on time. A quasi-2D trap has
$\lambda_z\gg\lambda_{x,y}$. For large anisotropies the condensate
shape along the z-direction is very similar to the Gaussian
profile of an ideal gas. However, along the weakly confined x- and
y-axes the condensate has a parabolic shape characteristic of the
hydrodynamic regime. The best description in terms of simple
analytic functions is therefore to model the condensate
wavefunction as a hybrid of parabola and Gaussian. Experiments on
the condensate expansion in various regimes show the smooth
crossover from hydrodynamic expansion to the characteristics of a
quasi-2D gas, which essentially expands like an ideal
non-interacting gas. To determine the dynamics of the quasi-2D
condensate we use a variational method, as introduced in
\cite{Perez-Garcia1996a}, and define the trial wavefunction
\begin{equation}
\psi=A_n\sqrt{1-\frac{x^2}{l_x^2}-\frac{y^2}{l_y^2}}e^{\frac{-z^2}{2l_z^2}}e^{\textrm{i}\left(\beta_x
x^2+\beta_y y^2+\beta_z z^2\right)},\label{trialhybrid}
\end{equation}
where the normalization constant $A_n$ is given by
\begin{equation}
A_n^2=\frac{2}{l_xl_yl_z\pi^{3/2}}.\label{normalization}
\end{equation}
The condensate width $l_i(t)$ and phase $\beta_i(t)$ parameters
are functions of time and their time evolution completely
describes that of the condensate. The condensate density profile
is at all times restricted to a parabolic shape in the radial
plane and a Gaussian shape along the highly compressed axial
direction. The Lagrangian density for the nonlinear
Schr\"{o}dinger equation is given by
\begin{eqnarray}
{\cal L}  &\equiv& \frac{1}{2}\textrm{i}\hbar \left(
\frac{\partial\psi^{\ast}}{\partial t}\psi -
\psi^{\ast}\frac{\partial\psi}{\partial t} \right) +
\frac{\hbar^{2}}{2m}
\left|{\bf \nabla}\psi\right|^{2}\nonumber\\
&&+ V_{{\rm ext}}\left({\bf r},t\right)\left|\psi\right|^{2} +
\frac{1}{2}gN\left|\psi\right|^{4}; \label{Lagrangian}
\end{eqnarray}
with the nonlinearity parameter $g=4\pi\hbar^2 a/m$, where $a$ is
the scattering length, $N$ is the number of atoms in the
condensate and $m$ is the atomic mass. After inserting the trial
wave-function (\ref{trialhybrid}) into Eq.(\ref{Lagrangian}) the
corresponding Lagrangian is found through integration $L=\int{\cal
L}~d^3x$; the four terms of Eq.(\ref{Lagrangian}) lead to
\begin{eqnarray}
&&L=
L_1+L_2+L_3+L_4=\nonumber\\
&&\frac{\hbar}{2}\left(\frac{\dot{\beta}_x
l_x^2}{3}+\frac{\dot{\beta}_yl_y^2}{3}+\dot{\beta}_z l_z^2\right)+\nonumber\\
&&\frac{\hbar^2}{m}\left(\frac{\beta_x^2 l_x^2}{3}+\frac{\beta_y^2
l_y^2}{3}+\beta_z^2 l_z^2+\frac{1}{4l_z^2}\right)+\nonumber\\
&&\frac{m}{4}\left(\frac{\omega_x^2l_x^2}{3}+
\frac{\omega_y^2l_y^2}{3}+\omega_z^2l_z^2\right)+ \nonumber\\&&
\frac{\sqrt{2} g N}{3l_xl_yl_z\pi^{3/2}},\label{Lagexpr}
\end{eqnarray}
where we omitted the `quantum pressure' term \cite{Stringari1996a}
for the $x$ and $y$ directions (where this term is divergent due
to the sharp boundaries of the condensate wavefunction in the
hydrodynamic regime) but retained it for the z-direction where the
condensate assumes the Gaussian shape of an ideal non-interacting
gas (as the term proportional to $1/l_z^2$). The quantum pressure
term is crucial in describing the dynamics. The total energy per
particle $E_{tot}$ and the chemical potential $\mu$ are given by
\begin{equation}
E_{tot}=E_{\text{kin}}+E_{\text{pot}}+E_{\text{int}}~~~~~~~\mu=E_{\text{kin}}+E_{\text{pot}}+2E_{\text{int}},\label{thermo}
\end{equation}
where $E_{\text{kin}},~E_{\text{pot}}$ and $E_{\text{int}}$ are
the kinetic, potential and interaction energy, given by the last
three terms of the Lagrangian (\ref{Lagexpr}), respectively. The
Euler Lagrange equations
\begin{equation}
\frac{d}{dt}\frac{\partial L}{\partial\dot{l}_i}=\frac{\partial
L}{\partial l_i},~~~~~~~~\frac{d}{dt}\frac{\partial
L}{\partial\dot{\beta}_i}=\frac{\partial L}{\partial
\beta_i}.\label{eulerlagrange1}
\end{equation}
yield the dynamic equations for the condensate widths $l_i$ and
phases $\beta_i$. We find for the widths
\begin{equation}
\dot{l}_i=\frac{2\hbar}{m}\beta_il_i.\label{dotequation}
\end{equation}
After differentiating Eqs. (\ref{dotequation}) once more w.r.t.
time one can express the resulting second order equation in terms
of the $l_i$ alone:
\begin{equation}
\ddot{l}_i=-\omega_i^2(t)l_i+\left(\frac{2}{\pi}\right)^{\frac{3}{2}}\frac{gN}{m}\frac{1}{l_il_xl_yl_z}
\left(1-\frac{2}{3}\delta_{iz}\right)+\frac{\hbar^2}{m^2}\frac{1}{l_i^3}\delta_{iz},\label{hybrid}
\end{equation}
where $\delta_{iz}=1$ for $i=z$ and $0$ otherwise. It is
convenient to express the above equation in dimensionless
quantities, so we introduce the dimensionless time $\tau$ and
widths $d_i$ defined by
\begin{equation}
d_i=\frac{l_i}{a_{0}},~~~~\tau =
t\omega_0\label{hybriddimensionlessunits},
\end{equation}
where $a_0=\sqrt{\hbar/(m\omega_0)}$ is the harmonic oscillator
length. In terms of these quantities
Eq.(\ref{hybriddimensionlessunits}) can be rewritten as
\begin{eqnarray}
\ddot{d}_i=-\lambda_i^2(t)d_i+\frac{C_p}{d_id_xd_yd_z}\left(1-\frac{2}{3}\delta_{iz}\right)+\frac{1}{d_z^3}\delta_{iz},
\label{hybriddimless}
\end{eqnarray}
where the constant $C_p=8\sqrt{\frac{2}{\pi}}\frac{a}{a_0}N$. To
find the ground state of Eqs.(\ref{hybriddimless}) we have to set
the left side equal to zero and solve the remaining coupled
nonlinear equations:
\begin{equation}
d_{i0}^2\lambda_{i0}^2=\frac{C_p}{d_{x0}d_{y0}d_{z0}}\left(1-\frac{2}{3}\delta_{iz}\right)+\frac{1}{d_{z0}^2}\delta_{iz}.\label{hybridground}
\end{equation}
This cannot be done analytically but it is straight forward to
find a numerical solution. After some algebra and using various
symmetries the three coupled equations can be reduced to one
polynomial equation. Introducing new dimensionless units $D_i$,
defined as the condensate widths $l_i$ normalized by the axial
harmonic oscillator length $a_z$, i.e. $D_i=l_i/a_z$, the
polynomial equation can be written as
\begin{equation}
\gamma^8=\frac{1}{3}\left(\frac{C_p\lambda_{x0}\lambda_{y0}}{\lambda_z^{3/2}}\right)^{1/2}\gamma^3+1,
\label{eightorderequation}
\end{equation}
where $D_{z}=\gamma^2$. There is only one real and positive
solution to this equation For the x- and y widths we find
\begin{equation}
D_{x}=\left(\frac{C_p}{D_{z}}\frac{\lambda_{y0}\lambda_{z0}^{5/2}}{\lambda_{x0}^3}\right)^{1/4},~~~~D_{y}=D_{x}\frac{\lambda_{x0}}{\lambda_{y0}}.\label{hybridradialwidth}
\end{equation}

\section{Criteria for quasi-2D} We shall now examine the case
where the anisotropy becomes very large. A solution to
Eq.(\ref{eightorderequation}) is then given by neglecting the
first term on the r.h.s. and solving the remaining equation. We
find that $\gamma^2=D_{z}=1$ and thus the approximate solution is
given by the axial harmonic oscillator length
\begin{equation}
l_{z0}=\sqrt{\frac{\hbar}{m\omega_{z}}}.\label{zideal}
\end{equation}
It is the minimum width the condensate shape can attain and it is
also the solution for the width of a non-interacting gas. For this
reason the gas along the z-direction is said to have the
characteristics of an ideal non-interacting gas. We will see that
this also applies to the expansion and the collective excitations
of the gas which become identical to those of an ideal gas in the
limit of large anisotropies.

It is interesting to examine the range of validity of this
approximation and find an estimate of the error. Demanding that
the first term on the r.h.s. of Eq.(\ref{eightorderequation}) is
much smaller than the second we see that the error or the
deviation from the ideal gas solution scales with the ratio of
$N\lambda_{x0}\lambda_{y0}/\lambda_{z0}^{3/2}$. The 2D-regime can
be reached by either decreasing the number or increasing the axial
frequency.

Now we calculate the chemical potential from Eq.(\ref{thermo}) and
the terms of the Lagrangian (\ref{Lagexpr}) and obtain after some
algebra
\begin{equation}
\tilde{\mu}=\frac{1}{2}m\omega_x^2l_{x0}^2,
~~~~\tilde{\mu}=\mu-\frac{\hbar\omega_{z}}{2}\label{chempot},
\end{equation}
where we used $\omega_x^2l_{x0}^2=\omega_y^2l_{y0}^2$, the
expression for $l_{z0}$ (Eq. \ref{zideal}) and other symmetries of
Eqs.(\ref{hybridground}). We find that the relation
$l_i=\sqrt{2\tilde{\mu}/m\omega_i^2},~~i=x,y$ is similar to that
of a hydrodynamic gas \cite{Dalfovo1999a}, only that for the
quasi-2D gas we use the chemical potential shifted by an amount
$\hbar\omega_z/2$ to calculate the radial width. Inserting
solution (\ref{zideal}) for the axial width into Eq.
(\ref{hybridradialwidth}) we obtain explicit expressions for the
radial width
\begin{equation}
l_{x0}^2 = a_0^2\left(8\sqrt{\frac{2}{\pi}}N
\frac{a}{a_0}\frac{\lambda_{y0}\lambda_{z0}^{\frac{1}{2}}}{\lambda_{x0}^3}
\right)^{\frac{1}{2}}\label{xwidth}
\end{equation}
and, after substituting into Eq. (\ref{chempot}), for the chemical
potential
\begin{equation}
 \mu =
\frac{\hbar\omega_z}{2}\left[1+\left(8\sqrt{\frac{2}{\pi}}N
\frac{a}{a_0}\frac{\lambda_{x0}\lambda_{y0}}{\lambda_{z0}^{\frac{3}{2}}}
\right)^{\frac{1}{2}}\right].\label{chempotexplicit}
\end{equation}
This expression shows that the chemical potential tends towards
$\hbar\omega_z/2$, the harmonic oscillator ground state energy,
which is the energy per particle and also the chemical potential
of the ideal non-interacting gas. The deviation from this value is
small for a quasi-2D gas, and interestingly, given by the same
value as the correction to the axial condensate width of
Eq.(\ref{eightorderequation}). Note, however, that this small
deviation is vital for consistency. It is proportional to the
square of the radial condensate width (see Eq. \ref{chempot}). If
it was zero the radial condensate width would also be zero which
is neither possible nor self consistent.  The expression for the
Q2D chemical potential should be compared to that of a 3D
hydrodynamic gas $\mu_{3D}$ for which we obtain from
\cite{Dalfovo1999a} after some rearrangements
\begin{equation}
\mu_{3d} =\left(15 N
\frac{a}{a_0}\frac{\lambda_{x0}\lambda_{y0}}{\lambda_{z0}^{\frac{3}{2}}}
\right)^{\frac{2}{5}}.\label{chempothydrodynamic}
\end{equation}
We observe that this expression tends towards zero for
$N\rightarrow 0$ and the power law is also different from the Q2D
expression (\ref{chempotexplicit}). In previous papers
\cite{Gorlitz2001} the condition $\mu_{3D}<\hbar\omega_z$ was
listed as a criterion for Q2D as $\mu_{3D}$ is on the order of the
interaction energy at low densities. Similarly we can impose the
conditon $\mu<\hbar\omega_z$ on the expression of the Q2D chemical
potential of Eq. (\ref{chempotexplicit}) and we find for the
maximum number of atoms to achieve 2D for a given trap geometry
\begin{equation}
N<C\sqrt{\frac{\hbar}{m
a^2}}\sqrt{\frac{\omega_z^3}{\omega_x^2\omega_y^2}},
\end{equation}
where the constant $C=\sqrt{32/225}$ for $\mu_{3D}<\hbar\omega_z$
and $C=\sqrt{\pi/256}$ for $\mu<\hbar\omega_z$.

\begin{figure}
\includegraphics[width=\columnwidth]{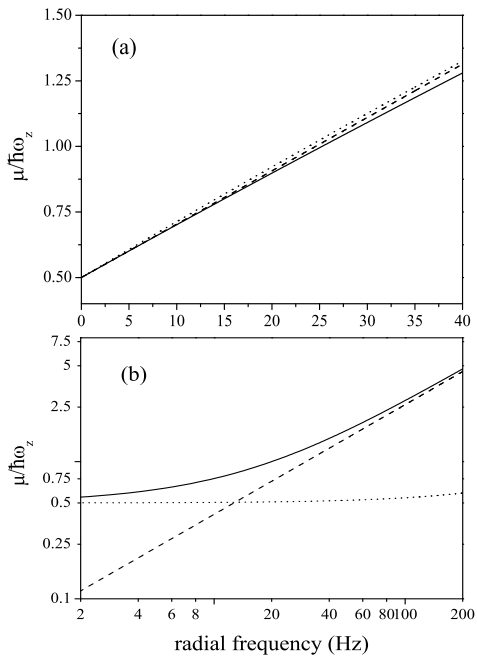}\\
\caption{(a) The chemical potential plotted in units of
$\hbar\omega_z$ against increasing radial trap frequency: hybrid
variational model (solid line), analytic approximation (dashed
line), Gaussian variational model (dotted line); (b) the chemical
potential plotted for a wider range of radial trap frequencies;
the ideal gas limit is given by the dotted line, the hydrodynamic
prediction by the dashed line.\label{plotchempot}}
\end{figure}

In Fig.~\ref{plotchempot}a we show plots of the chemical potential
as a function of increasing radial trap frequency
$\omega_r=\omega_x=\omega_y$, whereas the axial frequency remains
constant at $\omega_z/2\pi=2.2~\text{kHz}$ and the number of atoms
is taken to be $N=8\times 10^4$.
For very small values of $\omega_r$ the anisotropy
$A=\lambda_z/\lambda_x$ is very high and the chemical potential
approaches $\mu=\hbar\omega_z/2$. Gradually increasing the radial
trap frequency has the effect of reducing the anisotropy and
increasing the chemical potential. An exact numerical solution
using the variational hybrid model is given by the solid line. The
dashed line denotes the analytical approximation as given by Eq.
(\ref{chempotexplicit}) for a condensate well in the Q2D regime.
There is very good agreement between the two for frequencies up to
$\omega_r\simeq 20~\text{Hz}$, where the anisotropy is on the
order of $A\simeq 100$ and $\mu \approx \hbar\omega_{z}$. The
dotted line is given by the results of a Gaussian variational
model as described in \cite{Perez-Garcia1996a}, where the trial
wave-function in all spatial directions is given by a Gaussian. It
is remarkable how similar the results are for both models. In
physical reality the actual wavefunction in the radial direction
should be close to the inverted parabola we use in our `hybrid
model' (see Eq.\ref{trialhybrid}). The plot in Fig.
\ref{plotchempot}b shows the chemical potential for a wider range
of radial frequencies $\omega_r$ (solid line) on logarithmic
scales; this shows how the values bend from the hydrodynamic
asymptote (dashed line) to the ideal gas value of
$\hbar\omega_z/2$ (dotted line), with the bending point given by
values of $\omega_r/2\pi\approx 20~\text{Hz}$ and
$\mu\approx\hbar\omega_z$. From Eqs.
(\ref{Lagexpr},\ref{thermo},\ref{chempot}) we find the following
useful relations between the various energy contributions to the
Lagrangian:
\begin{equation}
E_{\text{kin}}=\frac{\hbar\omega_z}{4}=E_{\text{pot}}^z,~~~~
 E_{\text{pot}}^x=E_{\text{pot}}^y=\frac{E_{\text{int}}}{2}
\end{equation}
Using Eq.(\ref{thermo}) the total energy per particle is found
from the Lagrangian (\ref{Lagexpr}) to be
\begin{equation}
E_{tot}= \frac{\hbar\omega_z}{2}+\frac{m\omega_x^2l_{x0}^2}{3}=
\frac{\hbar\omega_z}{2}+\frac{2}{3}\tilde{\mu}.
\end{equation}
For large anisotropies and small atom numbers this expression
tends towards $\hbar\omega_z/2$, the harmonic oscillator ground
state energy. This is what we expected, given that in the Q2D
regime the gas is non-interacting and thus the energy per particle
is equal to the chemical potential. The release energy, defined as
the energy of the expanding cloud once the trap has been switched
off, is given by the sum of the in-trap kinetic and interaction
energies:
\begin{equation}
E_{\text{rel}}=E_{\text{kin}}+E_{\text{int}}=\frac{\hbar\omega_z}{4}+\frac{1}{3}\tilde{\mu}=\frac{E_{\text{tot}}}{2},
\end{equation}
which tends toward $\hbar\omega_z/4$, equal to half the ground
state energy, because the potential energy was lost when the atoms
were released from the trap. We also find the atomic peak density
$n_{\text{peak}}$ of the Q2D distribution, given by Eq.
(\ref{trialhybrid}), from Eqs.
(\ref{normalization},\ref{zideal},\ref{xwidth}):
\begin{equation}
n_{\text{peak}}=N A_n^2=\frac{\tilde{\mu}}{g}=
\left(\frac{N}{(2\pi)^{\frac{5}{2}}}
\frac{\lambda_{x0}\lambda_{y0}\lambda_{z0}^{\frac{1}{2}}}{a_0^5a}\right)^{\frac{1}{2}}.
\end{equation}

\section{Condensate expansion and release energy}
\begin{figure}
\includegraphics[width=7.5 cm]{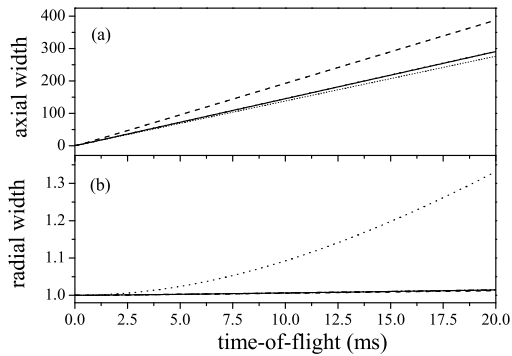}\\
\caption{The condensate axial (a) and radial (b) widths,
normalized by their initial values, plotted in TOF. The solid line
is based upon the hybrid model, the dotted upon the expansion of a
non-interacting wavepacket and the dashed line upon the
hydrodynamic theory.\label{expansionscaled}}
\end{figure}

It is interesting to examine the expansion of the condensate in
time-of-flight (TOF) in the 2D-regime. As a Q2D gas resembles an
ideal gas in the axial direction its expansion is described by the
Schr\"odinger equation and can be calculated analytically. Looking
at Eqs. (\ref{hybrid}), noting that $\lambda_z(t)=0$ for expansion
and neglecting the second term describing the interaction we
obtain $\ddot{l}_z=\hbar^2/(m^2l_z^3)$. The solution to this
second order differential equation for $l_z(0)=l_{z0}$ and
$\dot{l}_z=0$ is given by
\begin{equation}
l_z=l_{z0}\sqrt{1+\frac{\hbar^2t^2}{m^2l_{z0}^4}}=l_{z0}\sqrt{1+\omega_z^2t^2},\label{idealgasexpansion}
\end{equation}
where Eq. (\ref{zideal}) has been used. In the following we will
refer to this simple expression as the `ideal gas expansion'.
Fig.~\ref{expansionscaled} shows the axial and radial condensate
widths, normalized by their initial values (in the trap), during
TOF for a trap with $N=8\times 10^4$ atoms,
$\omega_z/2\pi=2.2~\text{kHz}$ and $\omega_r/2\pi=7~\text{Hz}$,
which are typical parameters for the experiment in
\cite{Smith2004b}. The prediction of the hybrid model, which is
indistinguishable from that of the Gaussian variational model, is
given by the solid line. For the axial direction, shown in (a), it
is very close to the expansion of the ideal gas given by Eq.
(\ref{idealgasexpansion}) and plotted as the dotted line. The
faster expansion of a hydrodynamic gas \cite{Castin1996a} is given
by the dashed line. These curves demonstrate clearly that for the
axial expansion the system behaves like an ideal gas. This
contrasts the radial expansion of Fig.~\ref{expansionscaled}b,
where the hybrid and Gaussian variational models (solid line)
agree nearly perfectly with the hydrodynamic theory (dashed line)
but are far removed from the ideal gas expansion (dotted line);
this demonstrates the hydrodynamic character of the radial
expansion.

\begin{figure}
\includegraphics[width=\columnwidth]{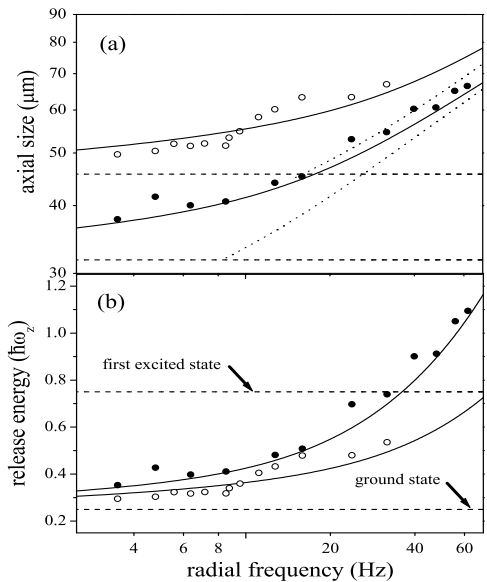}\\
\caption{(a) The axial expansion of the condensate in TOF for two
different trap geometries and atom numbers. Solid lines indicate
the theoretical predictions, dashed lines indicate the ideal gas
limit and dotted lines the hydrodynamic limit. The data are taken
for traps with initial oscillation frequencies (before release) of
$\omega_z/2\pi=1990~\text{Hz (open circles)},~960~\text{Hz (filled
circles)}$ \cite{Smith2004b}. The atom numbers are $8\times 10^4$
and $1.1\times 10^5$ for the upper and lower curves, respectively.
(b) release energies for the same data points; the dashed lines
indicate the energies of the ground and the first excited
state.\label{expansionvsradialfrequency}}
\end{figure}

In Fig.~\ref{expansionscaled} we plotted the condensate expansion
as a function of time; however, in the experiment in
\cite{Smith2004b} the expansion times are kept constant at
$15~\text{ms}$ and the radial trap frequency $\omega_r$ is varied
to explore the various regimes and the crossover to Q2D. The
measurements in \cite{Smith2004b} were done for two different
axial trap frequencies and atom numbers. The results of the
measurements are shown in Fig.~\ref{expansionvsradialfrequency}a,
where the open and filled circles are the data for traps in which
the atoms initially have axial oscillation frequencies of
$\omega_z/2\pi=1990~ \text{Hz}$ and $960~\text{Hz}$, respectively.
To obtain a theoretical comparison we propagate the hybrid model
Eqs. (\ref{hybriddimless}) repeatedly for varying initial
conditions obtained from Eq. (\ref{eightorderequation}). The
hybrid model predictions for the two optical traps are given by
the solid lines and we find good agreement to the experimental
data for both traps. The horizontal dashed lines indicate the
expansion of the ideal gas, given by Eq. (\ref{idealgasexpansion})
and the dotted lines the expansion of the hydrodynamic gas. The
`ideal gas' and hydrodynamic models yield straight lines on the
log-log plot. In contrast, the curve for the hybrid model follows
the hydrodynamic asymptote down to $\omega_r\approx20~\text{Hz}$,
corresponding to $\mu\approx\hbar\omega_z$, where it bends and
follows the 'ideal gas' line towards zero radial frequency. This
transition from the hydrodynamic to the `ideal gas' asymptote
gives conclusive evidence of the gas entering the Q2D regime.
There is a noticeable deviation for $\omega_r$ well below the
cross-over frequency because in the experiment imperfections broke
the radial symmetry and added a residual potential along the
y-axis. This was taken into account in the theoretical
calculations. There is good agreement with the experimental data.
The overall frequencies in x- and y-direction are given by
$\omega_x=\omega_r$ and
$\omega_y=\sqrt{\omega_r^2+(\omega_y^{\text{res}})^2}$. The
residual frequencies $\omega_y^{\text{res}}$ were measured in
\cite{Smith2004b} and are given by
$\omega_y^{\text{res}}/2\pi=26~\text{Hz}$ for the trap with
$\omega_z/2\pi=1990~\text{Hz}$ and
$\omega_y^{\text{res}}/2\pi=12~\text{Hz}$ for the trap with
$\omega_z/2\pi=960~\text{Hz}$.

It is also possible to calculate the release energy
$E_{\text{rel}}$ of the condensate from the experimental
measurements of its axial width as has been done in
\cite{Gorlitz2001}. The release energy is easily calculated after
a long time-of-flight. In this case the potential term in Eq.
(\ref{thermo}b) is zero and the interaction term negligible so
that we are left with the kinetic energy alone and find
$E_{\text{rel}}=E_{\text{kin}}$. The terms in x and y can be
neglected as most energy is in the previously tightly confined
z-direction and we obtain from the Lagrangian (\ref{Lagexpr}) an
expression for the release energy
\begin{equation}
E_{\text{rel}}\approx E_{\text{kin}}^z\approx
\frac{\hbar^2}{m}\left(l_z^2\beta_z^2+\frac{1}{4l_z^2}\right)=
\frac{m}{4}\dot{l}_z^2+\frac{1}{4}\frac{\hbar^2}{m}\frac{1}{l_z^2},\label{erelease}
\end{equation}
where we used relation (\ref{dotequation}) for the last step.
After an initial acceleration, when released from the trap, the
condensate moves with constant velocity and after long enough
time-of-flight (TOF) we can approximate $\dot{l}_z\approx l_z/t$,
where t is the TOF. The release energy is then written as
\begin{equation}
E_{\text{rel}}\approx\frac{m}{4}\frac{l_z^2}{t^2},\label{releaseenergy}
\end{equation}
where we neglected the quantum pressure proportional to $1/l_z^2$
as it tends to zero for long time-of-flights.
 In a Q2D gas the total energy is given by the sum of the
potential and the kinetic energy in the axial direction, both of
which contribute by an equal amount to the harmonic oscillator
ground state. As the potential energy is lost when the trap is
switched off the remaining kinetic energy constitutes half the
ground state energy so that $E_{\text{rel}}=\hbar\omega_z/4$. The
release energies for the same set of data as in (a) are plotted in
Fig. \ref{expansionvsradialfrequency}b, Eq.(\ref{releaseenergy})
was used to determine the release energy from the condensate
width. The theoretical predictions of the hybrid model are given
by the solid lines, the dashed lines indicate the value for the
ground state and the first excited state, respectively. In
\cite{Gorlitz2001} the expression $E_{\text{rel}}^p=m
L_z^2/(14t^2)$ is used to find the release energy, where $L_z$ is
the width of a fitted inverted parabola. This formula can be
derived by making a fully parabolic ansatz to the trial
wavefunction - an approach that is valid in the hydrodynamic limit
but is not really applicable in the Q2D limit, where one should
rather use Eq. (\ref{releaseenergy}) as otherwise the release
energy is underestimated  by about $20\%$.

There is excellent agreement between the theoretical predictions
and the experimental data and we find that the release energy does
indeed go towards $\hbar\omega_z/4$ indicating that the Q2D limit
was reached in the experiment \cite{Smith2004b}.

\section{Collective Excitations in Q2D} Another, as yet unexplored,
method to observe the transition to Q2D is to probe the collective
excitation spectrum. An ansatz for the trial wavefunction of the
type (\ref{trialhybrid}) allows for the description of the three
quadrupolar modes \cite{Dalfovo1999a}. In an axially symmetric
trap they are given by the $m=2$ mode and the $m=0$ low- and
high-lying modes, where $m$ denotes the angular momentum quantum
number. In order to calculate the mode frequencies we linearise
the dynamic equations of the hybrid model (\ref{hybriddimless})
around the ground state. Making the ansatz
\begin{equation}
d_i=d_{i0}+\epsilon_i,
\end{equation}
inserting it into Eqs.(\ref{hybriddimless}), expanding up to first
order in $\epsilon$, and using Eqs.(\ref{hybridground}) to
simplify and combine certain terms, we obtain after some algebra
\begin{equation}
\left(\begin{array}{c}
\ddot{\epsilon}_x\\
\ddot{\epsilon}_y\\
\ddot{\epsilon}_z
\end{array}\right)=
-\left(\begin{array}{ccc} \frac{3C_p}{d_{x0}^3d_{y0}d_{z0}} &
\frac{C_p}{d_{x0}^2d_{y0}^2d_{z0}} &
\frac{C_p}{d_{x0}^2d_{y0}d_{z0}^2} \\
\frac{C_p}{d_{x0}^2d_{y0}^2d_{z0}} &
\frac{3C_p}{d_{x0}d_{y0}^3d_{z0}} &
\frac{C_p}{d_{x0}d_{y0}^2d_{z0}^2} \\
\frac{C_p/3}{d_{x0}^2d_{y0}d_{z0}^2} &
\frac{C_p/3}{d_{x0}d_{y0}^2d_{z0}^2} &
\frac{C_p}{d_{x0}d_{y0}d_{z0}^3}+\frac{4}{d_{z0}^4}
\end{array}\right)
\left(\begin{array}{c}
\epsilon_x\\
\epsilon_y\\
\epsilon_z
\end{array}\right)\label{epsilonexpansion}
\end{equation}

\begin{figure}
\includegraphics[width=\columnwidth]{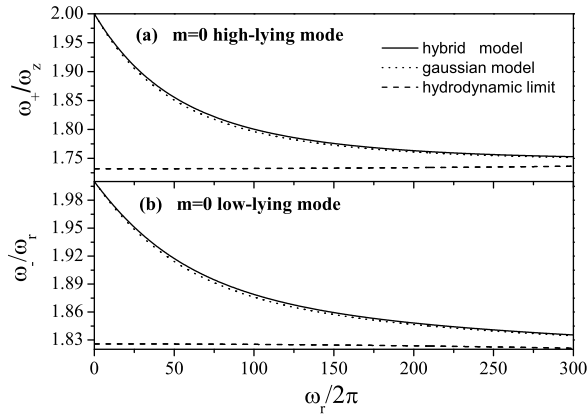}\\
\caption{The $m=0$ high-lying mode frequency ($\omega_+$) and the
$m=0$ low-lying mode frequency ($\omega_-$) plotted against the
radial trap frequency $\omega_r/2\pi$ in figures (a) and (b),
respectively; hybrid variational model (solid line), Gaussian
variational model (dotted line) and hydrodynamic prediction
(dashed line).\label{collfreqs}}
\end{figure}

Calculating the eigenvalues and eigenfrequencies of the above
matrix one finds the collective excitation frequencies and modes.
This can be easily done numerically. For simplicity that lends
itself to an easy analytical treatment we assume cylindrical
symmetry
$\lambda_{x0}=\lambda_{y0}\equiv\lambda_{r0},~d_{x0}=d_{y0}\equiv
d_{r0}$ and reduce the set of three equations to a set of two
\begin{equation}
\left(\begin{array}{c}
\ddot{\epsilon}_r\\
\ddot{\epsilon}_z
\end{array}\right)=-\left(\begin{array}{ll}
4 & D d_{z0}\\
\frac{2}{3}Dd_{z0} & 4\lambda_{z0}^2-\frac{1}{3}Dd_{r0}
\end{array}\right)
\left(\begin{array}{c}
\epsilon_r\\
\epsilon_z\\
\end{array}\right),\label{epsilon2d}
\end{equation}
where $D=C_p/(d_{r0}^3d_{z0}^3)$. We can diagonalize the above
matrix and find the eigenvalues which yields for the
eigenfrequencies $\omega^2$:
\begin{eqnarray}
&&\frac{\omega^2}{\omega_0^2}=
2\lambda_{z0}^2-\frac{1}{6}Dd_{r0}+2\nonumber\\&& \pm
\sqrt{\left(-2\lambda_{z0}^2+\frac{1}{6}Dd_{r0}+2\right)^2+\frac{2}{3}D^2d_{z0}^2}.
\label{mnodmodeshybrid}
\end{eqnarray}
Inserting expression (\ref{zideal},\ref{xwidth}) for the ground
state widths into the equation above we obtain an analytic
expression for the collective excitation frequencies in the Q2D
regime:
\begin{eqnarray}
&&\frac{\omega^2}{\omega_0^2}=2\lambda_{z0}^2-\frac{1}{6}\sqrt{\frac{C_p\lambda_{r0}^2}{\lambda_{z0}^{\frac{5}{2}}}}+2\nonumber\\
&& \pm
\sqrt{\left(-2\lambda_{z0}^2+\frac{1}{6}\sqrt{\frac{C_p\lambda_{r0}^2}{\lambda_{z0}^{\frac{5}{2}}}}+2\right)^2
+\frac{2}{3}\sqrt{C_p\lambda_{r0}^6\lambda_{z0}^{\frac{5}{2}}}}.\nonumber
\end{eqnarray}
 The two frequencies given by
Eq.(\ref{mnodmodeshybrid}) describe the  high- and low-lying $m=0$
modes of the collective excitation spectrum. The high-lying $m=0$
mode is an in-phase compressional mode along all directions
(breathing mode). The low-lying $m=0$ corresponds to a radial
oscillation of the width which is out of phase with an oscillation
along the trap axis. The third mode (not described by Eq.
\ref{mnodmodeshybrid}) in an axially symmetric trap is the $m=2$
mode. It corresponds to a quadrupole type excitation in the radial
plane and its frequency is given $\omega_2=\sqrt{2}\omega_r$,
irrespective of the axial frequency. Fig.~\ref{collfreqs} shows
the change of the collective excitation frequencies of the two
$m=0$ modes for increasing $\omega_r$. We find that the high-lying
mode frequency ($\omega_+$) changes from a value of
$\omega_+=2~\omega_z$ for radial frequencies close to zero to a
value of $\omega_+=\sqrt{3}~\omega_z$ for large radial
frequencies. The latter value has been determined from the
hydrodynamic model \cite{Dalfovo1999a}, given by the dashed line,
in the limit of $\omega_r\rightarrow 0$. The hardly
distinguishable predictions of the hybrid- and Gaussian
variational models are given by the solid and dotted lines,
respectively. For very small $\omega_r$, in the Q2D regime, the
mode frequencies approach the ideal or non-interacting gas limit
where the mode frequency is twice the trap frequency. In the ideal
gas limit the radial oscillation goes towards zero and we obtain a
pure axial oscillation for the high-lying mode, which can be found
from the eigenvectors of (\ref{epsilonexpansion}).

Something similar happens for the low-lying mode frequency
$\omega_-$ which changes from the hydrodynamic limit of
$\omega_-=\sqrt{10/3}~\omega_r$ to $\omega_-=2\omega_r$ for
decreasing $\omega_r$. An analysis of the eigenvector
(corresponding to this mode) of matrix (\ref{epsilonexpansion})
shows that in the Q2D regime the axial component of the
oscillation is increasingly suppressed and goes towards zero in
the limit of infinitely small $\omega_r$. Surprisingly, in this
limit the oscillation in the radial direction at the mode
frequency $\omega_-=2\omega_r$ seems not at all affected by the
hydrodynamic character and strong interactions of the gas in the
radial plane. It oscillates at the same frequency as an ideal
collisionless gas, although it is far from being collisionless.
This feature has been pointed out by Pitaevskii and Rosch
\cite{Pitaevskii1997a}, and depends on the special symmetry in the
two-dimensional regime.
\section{Conclusions and outlook}
We studied the properties of Q2D condensates using a hybrid
variational model based on a Gaussian-parabolic trial
wavefunction. The chemical potential and ground state energy were
calculated for a wide range of parameters from the hydrodynamic to
the Q2D regime. We find that the chemical potential approaches the
harmonic oscillator ground state and derive an analytical
expression for its values in the Q2D regime. The condensate size
in time-of-flight and its release energy are plotted for traps
with varying anisotropy. We find good agreement to recent
experimental results \cite{Smith2004b}. We also calculate the
excitation spectrum of the quadrupole modes of the Q2D gas and
find a gradual change from the hydrodynamic values to values equal
to twice the trap frequencies in the Q2D regime, as predicted in
\cite{Pitaevskii1997a}. All results from the hybrid model are
compared to those derived from a Gaussian variational model
\cite{Perez-Garcia1996a}, a hydrodynamic model \cite{Castin1996a}
and the ideal gas model. In future work we want to explore the
effect Q2D has on vortex structure, dynamics and decay.
\begin{acknowledgments}
The authors would like to acknowledge financial support from the
EPSRC and DARPA.
\end{acknowledgments}

\bibliography{intro}

\begin{thebibliography}{13}
\expandafter\ifx\csname natexlab\endcsname\relax\def\natexlab#1{#1}\fi
\expandafter\ifx\csname bibnamefont\endcsname\relax
  \def\bibnamefont#1{#1}\fi
\expandafter\ifx\csname bibfnamefont\endcsname\relax
  \def\bibfnamefont#1{#1}\fi
\expandafter\ifx\csname citenamefont\endcsname\relax
  \def\citenamefont#1{#1}\fi
\expandafter\ifx\csname url\endcsname\relax
  \def\url#1{\texttt{#1}}\fi
\expandafter\ifx\csname urlprefix\endcsname\relax\def\urlprefix{URL }\fi
\providecommand{\bibinfo}[2]{#2}
\providecommand{\eprint}[2][]{\url{#2}}

\bibitem[{\citenamefont{Paredes et~al.}(2004)\citenamefont{Paredes, Widera,
  Murg, Mandel, Folling, Cirac, Shlyapnikov, Hansch, and Bloch}}]{Paredes2004a}
\bibinfo{author}{\bibfnamefont{B.}~\bibnamefont{Paredes}},
  \bibinfo{author}{\bibfnamefont{A.}~\bibnamefont{Widera}},
  \bibinfo{author}{\bibfnamefont{V.}~\bibnamefont{Murg}},
  \bibinfo{author}{\bibfnamefont{O.}~\bibnamefont{Mandel}},
  \bibinfo{author}{\bibfnamefont{S.}~\bibnamefont{Folling}},
  \bibinfo{author}{\bibfnamefont{I.}~\bibnamefont{Cirac}},
  \bibinfo{author}{\bibfnamefont{G.}~\bibnamefont{Shlyapnikov}},
  \bibinfo{author}{\bibfnamefont{T.}~\bibnamefont{Hansch}}, \bibnamefont{and}
  \bibinfo{author}{\bibfnamefont{I.}~\bibnamefont{Bloch}},
  \bibinfo{journal}{NATURE} \textbf{\bibinfo{volume}{429}},
  \bibinfo{pages}{277} (\bibinfo{year}{2004}).

\bibitem[{\citenamefont{Gorlitz et~al.}(2001)\citenamefont{Gorlitz, Vogels,
  Leanhardt, Raman, Gustavson, Abo-Shaeer, Chikkatur, Gupta, Inouye, Rosenband
  et~al.}}]{Gorlitz2001}
\bibinfo{author}{\bibfnamefont{A.}~\bibnamefont{Gorlitz}},
  \bibinfo{author}{\bibfnamefont{J.}~\bibnamefont{Vogels}},
  \bibinfo{author}{\bibfnamefont{A.}~\bibnamefont{Leanhardt}},
  \bibinfo{author}{\bibfnamefont{C.}~\bibnamefont{Raman}},
  \bibinfo{author}{\bibfnamefont{T.}~\bibnamefont{Gustavson}},
  \bibinfo{author}{\bibfnamefont{J.}~\bibnamefont{Abo-Shaeer}},
  \bibinfo{author}{\bibfnamefont{A.}~\bibnamefont{Chikkatur}},
  \bibinfo{author}{\bibfnamefont{S.}~\bibnamefont{Gupta}},
  \bibinfo{author}{\bibfnamefont{S.}~\bibnamefont{Inouye}},
  \bibinfo{author}{\bibfnamefont{T.}~\bibnamefont{Rosenband}},
  \bibnamefont{et~al.}, \bibinfo{journal}{Phys. Rev. Lett.}
  \textbf{\bibinfo{volume}{87}} (\bibinfo{year}{2001}).

\bibitem[{\citenamefont{Rychtarik et~al.}(2004)\citenamefont{Rychtarik,
  Engeser, Nagerl, and Grimm}}]{Rychtarik2004}
\bibinfo{author}{\bibfnamefont{D.}~\bibnamefont{Rychtarik}},
  \bibinfo{author}{\bibfnamefont{B.}~\bibnamefont{Engeser}},
  \bibinfo{author}{\bibfnamefont{H.-C.} \bibnamefont{Nagerl}},
  \bibnamefont{and} \bibinfo{author}{\bibfnamefont{R.}~\bibnamefont{Grimm}},
  \bibinfo{journal}{Phys. Rev. Lett.} \textbf{\bibinfo{volume}{92}}
  (\bibinfo{year}{2004}).

\bibitem[{\citenamefont{Smith et~al.}(2004)\citenamefont{Smith, Heathcote,
  Hechenblaikner, Nugent, and Foot}}]{Smith2004b}
\bibinfo{author}{\bibfnamefont{N.~L.} \bibnamefont{Smith}},
  \bibinfo{author}{\bibfnamefont{W.}~\bibnamefont{Heathcote}},
  \bibinfo{author}{\bibfnamefont{G.}~\bibnamefont{Hechenblaikner}},
  \bibinfo{author}{\bibfnamefont{E.}~\bibnamefont{Nugent}}, \bibnamefont{and}
  \bibinfo{author}{\bibfnamefont{C.~J.} \bibnamefont{Foot}},
  \bibinfo{journal}{cond-mat/0410101}  (\bibinfo{year}{2004}).

\bibitem[{\citenamefont{Bishop and Reppy}(1978)}]{Bishop1978}
\bibinfo{author}{\bibfnamefont{D.}~\bibnamefont{Bishop}} \bibnamefont{and}
  \bibinfo{author}{\bibfnamefont{D.}~\bibnamefont{Reppy}},
  \bibinfo{journal}{Phys. Rev. Lett.} \textbf{\bibinfo{volume}{40}},
  \bibinfo{pages}{1727} (\bibinfo{year}{1978}).

\bibitem[{\citenamefont{Kosterlitz and Thouless}(1973)}]{Kosterlitz1973}
\bibinfo{author}{\bibfnamefont{J.}~\bibnamefont{Kosterlitz}} \bibnamefont{and}
  \bibinfo{author}{\bibfnamefont{D.}~\bibnamefont{Thouless}},
  \bibinfo{journal}{J. Phys. C} \textbf{\bibinfo{volume}{6}},
  \bibinfo{pages}{1181} (\bibinfo{year}{1973}).

\bibitem[{\citenamefont{Safonov et~al.}(1998)\citenamefont{Safonov, Vasilyev,
  Yasnikov, Lukashevich, and Jaakkola}}]{Safonov1998}
\bibinfo{author}{\bibfnamefont{A.}~\bibnamefont{Safonov}},
  \bibinfo{author}{\bibfnamefont{S.}~\bibnamefont{Vasilyev}},
  \bibinfo{author}{\bibfnamefont{I.}~\bibnamefont{Yasnikov}},
  \bibinfo{author}{\bibfnamefont{I.}~\bibnamefont{Lukashevich}},
  \bibnamefont{and} \bibinfo{author}{\bibfnamefont{S.}~\bibnamefont{Jaakkola}},
  \bibinfo{journal}{Phys. Rev. Lett.} \textbf{\bibinfo{volume}{81}},
  \bibinfo{pages}{4545} (\bibinfo{year}{1998}).

\bibitem[{\citenamefont{Das}(2002)}]{Das2002a}
\bibinfo{author}{\bibfnamefont{K.~K.} \bibnamefont{Das}},
  \bibinfo{journal}{Phys. Rev. A} \textbf{\bibinfo{volume}{66}},
  \bibinfo{pages}{053612} (\bibinfo{year}{2002}).

\bibitem[{\citenamefont{Perez-Garcia et~al.}(1996)\citenamefont{Perez-Garcia,
  H.Michinel, Cirac, Lewenstein, and Zoller}}]{Perez-Garcia1996a}
\bibinfo{author}{\bibfnamefont{V.}~\bibnamefont{Perez-Garcia}},
  \bibinfo{author}{\bibnamefont{H.Michinel}},
  \bibinfo{author}{\bibfnamefont{J.}~\bibnamefont{Cirac}},
  \bibinfo{author}{\bibfnamefont{M.}~\bibnamefont{Lewenstein}},
  \bibnamefont{and} \bibinfo{author}{\bibfnamefont{P.}~\bibnamefont{Zoller}},
  \bibinfo{journal}{Phys. Rev. Lett.} \textbf{\bibinfo{volume}{77}},
  \bibinfo{pages}{5320} (\bibinfo{year}{1996}).

\bibitem[{\citenamefont{Stringari}(1996)}]{Stringari1996a}
\bibinfo{author}{\bibfnamefont{S.}~\bibnamefont{Stringari}},
  \bibinfo{journal}{Phys. Rev. Lett.} \textbf{\bibinfo{volume}{77}},
  \bibinfo{pages}{2360} (\bibinfo{year}{1996}).

\bibitem[{\citenamefont{Dalfovo et~al.}(1999)\citenamefont{Dalfovo, Giorgini,
  Pitaevskii, and Stringari}}]{Dalfovo1999a}
\bibinfo{author}{\bibfnamefont{F.}~\bibnamefont{Dalfovo}},
  \bibinfo{author}{\bibfnamefont{S.}~\bibnamefont{Giorgini}},
  \bibinfo{author}{\bibfnamefont{L.~P.} \bibnamefont{Pitaevskii}},
  \bibnamefont{and}
  \bibinfo{author}{\bibfnamefont{S.}~\bibnamefont{Stringari}},
  \bibinfo{journal}{Rev. Mod. Phys.} \textbf{\bibinfo{volume}{71}},
  \bibinfo{pages}{463} (\bibinfo{year}{1999}).

\bibitem[{\citenamefont{Castin and Dum}(1996)}]{Castin1996a}
\bibinfo{author}{\bibfnamefont{Y.}~\bibnamefont{Castin}} \bibnamefont{and}
  \bibinfo{author}{\bibfnamefont{R.}~\bibnamefont{Dum}},
  \bibinfo{journal}{Phys. Rev. Lett.} \textbf{\bibinfo{volume}{77}},
  \bibinfo{pages}{5315} (\bibinfo{year}{1996}).

\bibitem[{\citenamefont{Pitaevskii and Rosch}(1997)}]{Pitaevskii1997a}
\bibinfo{author}{\bibfnamefont{L.}~\bibnamefont{Pitaevskii}} \bibnamefont{and}
  \bibinfo{author}{\bibfnamefont{A.}~\bibnamefont{Rosch}},
  \bibinfo{journal}{Phys. Rev. A} \textbf{\bibinfo{volume}{55}},
  \bibinfo{pages}{R853} (\bibinfo{year}{1997}).

\end{thebibliography}

\end{document}